\documentclass[twocolumn,aps,prd,superscriptaddress,nofootinbib,amsmath,amssymb,floats,floatfix,nofrontpages,showkeys,longbibliography]{revtex4}
\usepackage[dvips]{graphicx,color}
\usepackage{subfig}
\usepackage{times}
\usepackage{mathrsfs}  
\usepackage{diagbox}
\usepackage{xcolor}
\usepackage[%
  colorlinks=true,
  urlcolor=blue,
  linkcolor=red,
  citecolor=blue
]{hyperref}
\usepackage{booktabs}
\usepackage{tabularx}
\usepackage{array}
\usepackage{threeparttable}
\RequirePackage{mathrsfs}
\RequirePackage{amsmath}
\RequirePackage{amssymb}
\RequirePackage{amsbsy}
\usepackage{graphicx}
\usepackage{multirow} 
\usepackage{caption}
\def\rf#1{(\ref{#1})}

\def\de#1/de#2{\frac{\partial {#1}}{\partial {#2}}}

\newcommand{\ba}{\begin{eqnarray}}
\newcommand{\ea}{\end{eqnarray}}
\newcommand{\be}{\begin{equation}}
\newcommand{\ee}{\end{equation}}

\usepackage{orcidlink}

\begin{document}

\title{Radial oscillations of pulsating neutron stars: The UCIa equation-of-state case}

\author{Grigoris Panotopoulos \orcidlink{0000-0003-1449-9108}} \email[Corresponding Author:]{grigorios.panotopoulos@ufrontera.cl}
\affiliation{Departamento de Ciencias F{\'i}sicas, Universidad de La Frontera, Casilla 54-D, 4811186 Temuco, Chile.}

\author{Ali \"Ovg\"un \orcidlink{0000-0002-9889-342X}}
\email{ali.ovgun@emu.edu.tr}
\affiliation{Physics Department, Eastern Mediterranean University, Famagusta, Cyprus.}

\author{Tamanna Iqbal \orcidlink{https://orcid.org/0000-0002-7442-2463}}
\email{iqbaltamanna11@gmail.com}
\affiliation{Department of Physics, Babasaheb Bhimrao Ambedkar University (A Central University), Lucknow 226025, India}

\author{Yashmitha Kumaran \orcidlink{https://orcid.org/0009-0006-5655-6931}}
\email{yashmithakmoses@gmail.com}
\affiliation{Instituto de Astronomía, Universidad Nacional Autónoma de México, AP 70-264, Ciudad de México 04510, México.}

\author{B.~K.~Sharma \orcidlink{https://orcid.org/0000-0003-0543-1582}}
\email{bk\_sharma@cb.amrita.edu}
\affiliation{Department of Physics, Amrita School of Physical Sciences, Amrita Vishwa Vidyapeetham, Coimbatore 641112, India.}

\begin{abstract}
Radial oscillations provide a clean dynamical test of the high-density stiffness of neutron-star equations of state. We study spherically symmetric pulsations of nonrotating relativistic stars built from cold, charge-neutral, $\beta$-equilibrated pure nucleonic matter described within relativistic mean-field theory. As a baseline we adopt the UCIa parameter set [Astron. Astro-
phys. 689, A242 (2024)], and we implement high-density stiffening via the $\sigma$-cut scheme by adding a regulator potential $U_{\rm cut}(\sigma)$ [Phys. Rev. C 92, no.5, 052801 (2015), Phys. Rev. C 106, no.5, 055806 (2022)]. For representative choices $f_s=0$ (no cutoff) and $f_s=0.58$ (stiffened), we solve the Tolman-Oppenheimer-Volkoff and tidal perturbation equations to obtain equilibrium sequences, mass-radius relations, and tidal deformabilities. We then derive and solve the linear general-relativistic radial pulsation equations to compute the eigenfrequencies and eigenfunctions of the fundamental and overtone modes. The $\sigma$-cutoff suppresses the growth of the scalar field at supranuclear density, increases the pressure, and shifts the maximum mass, radii, and $\Lambda_{1.4}$ accordingly, while systematically raising the radial-mode frequencies at fixed mass. Using the sign change of $\omega_0^2$ as a stability criterion, we identify stiffened models that remain radially stable up to the observed $\sim 2M_\odot$ mass scale and are consistent with current multimessenger constraints, demonstrating how radial spectra complement static EoS tests.
\end{abstract}


\maketitle

\section{Introduction}

Neutron stars are prime laboratories for cold ultradense matter: their interiors probe baryon densities from the outer crust-where nuclei coexist with a degenerate electron gas-to central densities that can reach several times nuclear saturation, $n_0 \simeq 0.16~\mathrm{fm}^{-3}$ \cite{LattimerPrakash:2007,OzelFreire:2016}. In this regime, the equation of state (EoS) of charge-neutral, $\beta$-equilibrated matter is still uncertain, yet it controls key macroscopic observables such as the maximum mass and radius of hydrostatic configurations, the tidal response in binaries, and dynamical properties encoded in oscillation spectra \cite{LattimerPrakash:2007,OzelFreire:2016}.

\smallskip

Observations have now carved out a nontrivial target space for dense-matter models. Precision pulsar timing has established neutron stars with masses close to $2M_\odot$, enforcing a robust lower bound on the maximum mass $M_{\max}$ supported by any viable EoS and favouring sufficiently stiff behaviour at supranuclear densities \cite{Demorest:2010,Antoniadis:2013,Cromartie:2019}. In parallel, X-ray pulse-profile modelling with NICER has provided radius constraints from hotspots on millisecond pulsars, including PSRJ0030+0451 \cite{Riley:2019,Miller:2019}, PSRJ0740+6620 \cite{Miller:2021qha,Riley:2021pdl}, and PSR~J0437-4715 \cite{Choudhury:2024xbk}. Gravitational-wave observations of binary neutron star mergers, most notably GW170817, constrain tidal deformabilities and thus the pressure at a few times $n_0$ \cite{Abbott:2018,Abbott:2019}. Taken together, these multimessenger inputs motivate EoSs that can satisfy the $2M_\odot$ requirement while remaining compatible with radius and tidal-deformability inferences \cite{OzelFreire:2016,Abbott:2019}.

\smallskip

From the modeling side, relativistic mean-field (RMF) theory provides a widely used microphysical framework that can be calibrated to nuclear properties near saturation and then extrapolated to the neutron-star core \cite{LattimerPrakash:2007,OzelFreire:2016}. In this context, a controlled way to stiffen the supranuclear sector without spoiling the near-$n_0$ calibration is the $\sigma$-cut scheme \cite{Maslov:2015,Patra:2022}. In this approach, a sigma cutoff potential $U_{\rm cut}(\sigma)$ is added to the scalar sector to curb the growth of the $\sigma$ field at large densities. Physically, this suppresses scalar attraction (effectively increasing baryon masses) and raises the pressure predominantly at high density, while leaving saturation properties essentially intact \cite{Maslov:2015,Patra:2022}. In the present work we employ the recently calibrated RMF parameterisation UCIa as baseline \cite{Malik:2024} and focus on cold, $\beta$-equilibrated \emph{pure nucleonic} matter, so that the impact of scalar-sector stiffening can be isolated cleanly.

\smallskip

Asteroseismology is a powerful and widely used technique that enables the study of the internal structure of stars, as the frequencies of oscillation modes in pulsating objects are highly sensitive to their composition and underlying equations-of-state. It has proven to be an essential tool for probing the interiors of the Sun, solar-like stars, and other stellar classes \cite{Turck-Chieze:2010rvs, Chaplin:2013hha}, providing information on nuclear reactions, chemical composition, equations of state, differential rotation, and meridional circulation.
More recently, Asteroseismology has been extended to the study of compact objects \cite{Kokkotas:1999bd, Paschalidis:2016vmz, DiClemente:2020szl, Kain:2020zjs}, including white dwarfs \cite{Corsico:2019nmr}, neutron stars \cite{Sagun:2020qvc}, strange quark stars \cite{Panotopoulos:2017eig}, dark matter stars \cite{Leung:2011zz, Leung:2012vea, Panotopoulos:2017eig}, and dark energy stars \cite{Panotopoulos:2020kgl}. For further details, the reader is referred to \cite{Lopes1, Lopes2, Kokkotas:2000up, Miniutti:2002bh, Passamonti:2004je, Passamonti:2005cz, Savonije:2007ay, VasquezFlores:2010eq, Brillante:2014lwa}.

\smallskip

Concerning excitation and detectability, several astrophysical mechanisms can excite stellar oscillation modes, including tidal interactions and resonant excitation in binary systems, mass accretion in low-mass X-ray binaries, starquakes induced by crustal fractures, supernova explosions, magnetic field reconfiguration, and other forms of instability \cite{Sagun:2020qvc, Franco:1999pma,kokkotas,tsang, Hinderer:2016eia, Chirenti:2016xys}. The Kepler and CoRoT missions have already measured oscillation spectra of solar-like stars, white dwarfs, and red giants \cite{obs1,obs2,obs3,obs4,obs5,obs6}. The limited sensitivity of current gravitational-wave detectors in the kHz frequency range prevents the detection of the radial oscillations considered here. However, third-generation ground-based gravitational-wave detectors, such as the Einstein Telescope \cite{ET} and the Cosmic Explorer \cite{CE}, are expected to achieve sensitivities approximately one order of magnitude higher than that of LIGO \cite{Sagun:2020qvc}. Such observations could enable simultaneous measurements of stellar masses, tidal Love numbers, oscillation frequencies, damping times, mode amplitudes, and moments of inertia.

\smallskip

Beyond static observables, the EoS controls the dynamics of relativistic stars. In particular, radial oscillations-spherically symmetric pulsations about hydrostatic equilibrium-provide a sharp diagnostic of dynamical stability in General Relativity \cite{Chandrasekhar:1964zz,Chandrasekhar:1964zza,Kokkotas:2000up}. The squared eigenfrequencies $\omega_n^2$ of the normal modes change sign at the onset of instability: a configuration becomes marginally stable when the fundamental mode satisfies $\omega_0^2=0$, while higher overtones encode additional information about the stiffness of the stellar fluid and how the equilibrium profile responds to changes in the EoS. Radial-mode calculations for relativistic stars have a long history \cite{Kokkotas:2000up} and remain useful as an EoS discriminator in modern contexts \cite{DiClemente:2020szl,Kain:2020zjs,Sagun:2020qvc}. Direct observational access to high-frequency radial spectra is challenging with current ground-based detectors; any near-term prospects rely on future improvements in sensitivity at kHz frequencies and on astrophysical excitation mechanisms \cite{Sagun:2020qvc,ET,CE}. This motivates using radial oscillations primarily as a \emph{theoretical} consistency check: they test whether a proposed high-density stiffening that satisfies static constraints also yields configurations that are dynamically stable.

\smallskip

A large fraction of EoS studies is therefore built around static structure and tidal observables alone. The central point of this paper is that, for stiffened RMF models, the radial spectrum provides an additional, independent filter: it ties the microphysical modification (here, scalar-sector regulation through $U_{\rm cut}(\sigma)$) directly to dynamical stability and mode systematics, rather than only to $(M,R,\Lambda)$. In this sense, radial oscillations provide a complementary consistency test for dense-matter modelling with high-density regulators.

\smallskip

In this work we: (i) present a transparent and self-contained derivation of the modified RMF field equations in the presence of the $\sigma$-cutoff potential $U_{\rm cut}(\sigma)$, clarifying how it alters the scalar response at high density \cite{Maslov:2015,Patra:2022}; (ii) construct cold, $\beta$-equilibrated pure-nucleonic EoSs based on UCIa \cite{Malik:2024} with representative cutoff choices $f_s=0$ and $f_s=0.58$; (iii) compute equilibrium sequences by solving the Tolman--Oppenheimer--Volkoff equations and determine tidal deformabilities in General Relativity; (iv) solve the linear radial pulsation equations to obtain eigenfrequencies and eigenfunctions for the fundamental and overtone modes, using the sign of $\omega_0^2$ to diagnose stability \cite{Chandrasekhar:1964zz,Kokkotas:2000up}; and (v) confront the resulting macroscopic and dynamical predictions with the multimessenger constraints that motivate stiff high-density behavior \cite{Demorest:2010,Antoniadis:2013,Cromartie:2019,Abbott:2018,Abbott:2019,Riley:2019,Miller:2019}.


\smallskip


\smallskip

The main goal of this paper is to provide a transparent and self-contained derivation of the modified RMF field
in the presence of the $\sigma$-cutoff potential $U_{\rm cut}(\sigma)$, emphasizing how the  $U_{\rm cut}(\sigma$)  
alters the scalar response and the high-density behavior of the EoS. We also aim to construct the equilibrium stellar 
sequences and compute their radial pulsation spectra in General Relativity, thereby assessing
the interplay between the $\sigma$-cutoff potential governed by parameter $f_s$, global observables 
$(M_{\max}, R_{1.4}, \Lambda_{1.4})$, and dynamical stability against radial oscillations.  


\smallskip

Our work is organised as follows: After this introductory section, we summarise the structure equations for interior solutions of fluid spheres in Section II, while the equation-of-state is discussed in detail in Section III. Next, in Section IV, we present the perturbation equations for pulsating stars, and in Section V we discuss the results of the present study. Finally, we conclude and summarise our work in Section VI. Throughout the manuscript we work in natural geometrical units, $\hbar=c=G=1$, and adopt the metric signature $(-,+,+,+)$. Unless stated otherwise, we consider non-rotating, spherically symmetric configurations governed by the Tolman- Oppenheimer-Volkoff (TOV) equations of General Relativity.


\section{Relativistic stars in the context of General Relativity}

We shall be working with fluid spheres made of isotropic matter in (1+3)-dimensional space-times assuming a vanishing cosmological constant. Although rotation may be present in astronomical objects, in the present work we shall consider non-rotating stars only. The inclusion of a non-vanishing angular momentum, at least in the slow-rotating case, may be implemented afterwards in a straightforward manner, see e.g. \cite{Staykov:2014mwa, Panotopoulos:2021dtu}.

We shall briefly review the structure equations for interior stellar solutions, starting from the field equations of Einstein's General Relativity \cite{Einstein:1915ca}
\begin{equation}
G_{mn} \equiv R_{mn} - \frac{1}{2} \: R \: g_{mn} = 8 \pi  T_{mn},
\end{equation}

\medskip

\noindent where $T_{mn}$ is the energy-momentum tensor of matter content, $g_{mn}$ is the metric tensor, $R_{mn}$ and $R$ are the Ricci tensor and Ricci scalar, respectively, while $G_{mn}$ is the Einstein tensor.

\smallskip

For static, spherically symmetric geometries the most general line element in Schwarzschild-like coordinates $\{ t, r, \theta, \phi \}$ is given by
\begin{equation}
    d s^2 = - e^{\nu} d t^2 + e^{\lambda} d r^2 + r^2 (d \theta^2 + \sin^2 \theta \: d \phi^2) , 
\end{equation}

\medskip

\noindent while the isotropic matter content, viewed as a perfect fluid, is described by a stress-energy tensor of the following form
\begin{equation}
T_a^b = Diag(-\rho, p, p, p),
\end{equation}

\medskip

\noindent where $p$ is the pressure of the fluid, in both radial and tangential directions, while $\rho$ is its energy density. What is more, depending on the matter content, $p$ and $\rho$ satisfy a certain EoS $F(p, \rho)=0$.

\smallskip

To compute the stellar mass, $M$, and radius, $R$, we need to integrate the Tolman-Oppenheimer-Volkoff (TOV) equations \cite{Oppenheimer:1939ne, Tolman:1939jz}. Those are the structure equations that permit us to obtain stellar interior solutions describing hydrostatic equilibrium. The TOV equations, assuming only one fluid component, are given by
\begin{align}
    \displaystyle m'(r) &= 4 \pi r^2 \rho (r) , \\ 
    \displaystyle \nu' (r) &= 2 \: \frac{m(r) + 4 \pi r^3 p(r)}{ r^2 \left( 1 - 2 m(r) / r \right) } , \\
    \displaystyle p'(r) &= - [ \rho(r) + p(r) ] \; \frac{\nu' (r)}{2} ,
\end{align}

\smallskip

\noindent where the prime denotes differentiation with respect to the radial coordinate $r$, while the mass function, $m(r)$, is defined as usual by
\begin{equation}
    \displaystyle e^{\lambda} = \frac{1}{1 - \frac{2 \: m(r)}{r}} .
    \label{eq:2}
\end{equation}

\medskip

In practice we only need to solve a system of two coupled differential equations
\begin{align}
    \displaystyle m'(r) &= 4 \pi r^2 \rho (r) , \\ 
    \displaystyle p'(r) &= - [ \rho(r) + p(r) ] \;  \frac{m(r) + 4 \pi r^3 p(r)}{ r^2 \left( 1 - 2 m(r) / r \right) } .
\end{align}

\smallskip

To find the solution, we integrate the structure equations throughout the star imposing initial conditions at the center ($r = 0$), and then we impose the matching conditions at the surface of the star ($r = R$). In particular, the initial conditions at the origin are as follows 
\begin{equation}
    m(0) = 0 ,
\end{equation}
\begin{equation}
    p(0) = p_c ,
\end{equation}

\medskip

\noindent with $p_{c}$ being the central pressure, while the exterior vacuum solution is given by Schwarzschild geometry \cite{Schwarzschild:1916uq}
\begin{align}
d s^2 = -f(r) d t^2 + f(r)^{-1} & dr^2 + r^2 (d \theta^2 + \sin^2 \theta \: d \phi^2), \\
f(r) &= 1- \frac{2M}{r} .\nonumber
\end{align}

Furthermore, the matching conditions yield
\begin{equation}
p(R) = 0,
\end{equation}
\begin{equation}
m(R) = M, 
\end{equation}
\begin{equation}
e^{\nu(R)} = 1-2 \frac{M}{R} .
\end{equation}

\medskip

The first two conditions allow us to compute the radius and the mass of the star. Finally, the other metric potential, $\nu(r)$, may be computed by
\begin{equation}
    \displaystyle \nu (r) = \ln \left( 1 - \frac{2 M}{R} \right) + 2 \int_R^r \frac{m(x) + 4 \pi x^3 p(x)}{ x^2 \left( 1 - 2 m(x) / x \right) } \: dx .
    \label{eq:3}
\end{equation}
where we have used the third matching condition.

\section{RMF Lagrangian with a $\sigma$-cutoff potential}

We begin by specifying the microscopic description of dense matter that underlies our stellar models. Within the RMF approach, 
nucleons interact via the exchange of isoscalar-scalar $\sigma$, isoscalar-vector $\omega_\mu$, and isovector-vector $\vec{\rho}_\mu$ 
mesons, not to be confused with the energy density of the matter content $\rho(r)$. Nonlinear self-interactions and mixed couplings among the mesons are introduced in order to reproduce empirical properties of 
finite nuclei and nuclear matter, and leptons ($e^-,\mu^-$) are included to ensure charge neutrality and $\beta$-equilibrium.

In the mean-field approximation for uniform matter, spatial gradients vanish and only the time components of the vector fields survive. 
The Lagrangian density we adopt can be written as \cite{Sumiyoshi:2019qgs,Thakur:2025axg}
\begin{align}
\mathcal{L} &= \bar{\Psi}\!\left[\gamma^\mu\!\left(i\partial_\mu - g_{\omega}\omega_\mu - 
g_{\rho}\,\vec{t}\!\cdot\!\vec{\rho}_\mu\right) - \left(m - g_{\sigma}\sigma\right)\right]\!\Psi \nonumber\\
&\quad + \frac{1}{2}\left(\partial_\mu\sigma\,\partial^\mu\sigma - m_\sigma^2\sigma^2\right) - 
\frac{1}{4}F^{(\omega)}_{\mu\nu}F^{(\omega)\mu\nu} - \frac{1}{4}\vec{F}^{(\rho)}_{\mu\nu}\!\cdot\!\vec{F}^{(\rho)\mu\nu} \nonumber\\
&\quad + \frac{1}{2}m_\omega^2\,\omega_\mu\omega^\mu + \frac{1}{2}m_\rho^2\,\vec{\rho}_\mu\!\cdot\!\vec{\rho}^\mu
 - \frac{1}{3}b\,m\,(g_{\sigma}\sigma)^3 - \frac{1}{4}c\,(g_{\sigma}\sigma)^4 \nonumber\\
&\quad + \frac{\xi}{4!}\,g_{\omega}^4\,(\omega_\mu\omega^\mu)^2
+ \Lambda_\omega\,g_{\rho}^2 g_{\omega}^2\,(\vec{\rho}_\mu\!\cdot\!\vec{\rho}^\mu)\,(\omega_\nu\omega^\nu)
- U_{\rm cut}(\sigma)
\nonumber\\
&\quad + \sum_{\ell=e,\mu}\bar{\Psi}_\ell(i\gamma^\mu\partial_\mu - m_\ell)\Psi_\ell,
\label{eq:Lagrangian}
\end{align}
where $F^{(\omega)}_{\mu\nu}=\partial_\mu\omega_\nu-\partial_\nu\omega_\mu$ and 
$\vec{F}^{(\rho)}_{\mu\nu}=\partial_\mu\vec{\rho}_\nu-\partial_\nu\vec{\rho}_\mu$ are the field-strength tensors 
of the vector mesons.

The first line of Eq.~\rf{eq:Lagrangian} describes nucleons moving in classical meson fields. The scalar field $\sigma$ generates 
an attractive potential and reduces the Dirac effective mass $m^\star$, while the $\omega_\mu$ and $\vec{\rho}_\mu$ fields provide 
repulsive isoscalar and isovector contributions, respectively. The cubic and quartic self-interactions of $\sigma$ controlled by 
$b$ and $c$ are crucial to reproduce nuclear matter saturation, whereas the quartic vector term and the mixed $\omega$-$\rho$ coupling, 
proportional to $\xi$ and $\Lambda_\omega$, allow one to tune the high-density behavior and the density dependence of the 
symmetry energy. The term $U_{\rm cut}(\sigma)$ is the additional high-density regulator whose role will be clarified below.

The sigma-cutoff potential ($U_{\rm cut}(\sigma)$) is given by  \cite{Zhang:2018lpl,Maslov:2015,Thakur:2025axg}
\begin{equation}
U_{\rm cut}(\sigma)=\alpha\,\ln\!\left[1+\exp\!\left\{\beta\!\left(\frac{g_{\sigma N}\sigma}{m_N}-f_s\right)\!\right\}\right],
\qquad
\label{eq:Ucut}
\end{equation}
where $\alpha$= $m_\pi^4$ and $\beta$= 120 to make EoS stiffer at high density. The dimensionless parameter $f_s$ is 
a free parameter which is 0.00 for original UCIa parameter set. We take $f_{s}$= 0.00 and 0.58 for the present study.    

\subsubsection*{Relativistic mean-field equations and effective masses}

In uniform matter the spatial components of the vector fields vanish, $\omega_i=\rho_i=0$, and the meson fields are determined 
by stationarity of the energy density with respect to $\sigma$, $\omega$ and $\rho$. Introducing the Dirac effective mass
\begin{equation}
m^\star = m - g_{\sigma}\sigma,
\end{equation}

The meson field equations can be written in a compact form as \cite{Malik:2024}
\begin{align}
\sigma &= \frac{g_{\sigma}}{m_{\sigma,\mathrm{eff}}^2}\sum_{i=n,p} n_{S,i}, \nonumber
\\
\omega &= \frac{g_{\omega}}{m_{\omega,\mathrm{eff}}^2}\sum_{i=n,p}n_{i} \label{eq:fields}
\\ \rho &= \frac{g_{\rho}}{m_{\rho,\mathrm{eff}}^2}\sum_{i=n,p} I_{3}n_{i} \nonumber
\end{align}
where $n_{S,i}$ and $n_{i}$ represent the scalar density and the number density of nucleon $i$,
respectively, and 
\begin{align}
m_{\sigma,\mathrm{eff}}^2 &= m_\sigma^2 + b\,m g_{\sigma}^3\sigma + c\,g_{\sigma}^4\sigma^2 + \frac{U'_{\rm cut}(\sigma)}{\sigma},
\label{eq:msigeff}\\
m_{\omega,\mathrm{eff}}^2 &= m_\omega^2 + \frac{\xi}{3!}\,g_{\omega}^4\,\omega^2 + 2\Lambda_\omega\,g_{\rho}^2 g_{\omega}^2\,\rho^2,
\label{eq:momeff}\\
m_{\rho,\mathrm{eff}}^2 &= m_\rho^2 + 2\Lambda_\omega\,g_{\rho}^2 g_{\omega}^2\,\omega^2.
\label{eq:mrhoeff}
\end{align}

Differentiating Eq.~\rf{eq:Ucut} with respect to $\sigma$ we obtain
\begin{equation}
U'_{\rm cut}(\sigma)=\alpha\,\beta\,\frac{g_{\sigma}}{m}\,
\frac{1}{1+\exp\!\left[-\beta\!\left(\frac{g_{\sigma}\sigma}{m}-f_s\right)\right]},
\label{eq:UcutPrime}
\end{equation}

\subsubsection*{$\beta$-equilibrium and charge neutrality}

We now impose the conditions of cold catalyzed matter relevant for neutron stars. The chemical potentials of 
nucleons and leptons in the RMF mean-field background are given by
\begin{align}
\mu_i &= \sqrt{k_{F,i}^2+m^{\star 2}}+g_{\omega}\omega+g_{\rho}I_{3}\rho, \qquad (i= n,p)\\
\mu_\ell &= \sqrt{k_{F,\ell}^2+m_\ell^2}, \qquad (\ell=e,\mu).
\end{align}
The first term in $\mu_i$ represents the Fermi energy of nucleons with effective mass $m^\star$, while the remaining pieces 
correspond to the vector mean-field shifts.
The $\beta$-equilibrium and electric charge neutrality were imposed, which are respectively defined by the relations
\begin{align}
\mu_n &= \mu_p + \mu_e, \qquad
\mu_\mu = \mu_e,
\qquad
n_{p}= n_{e} + n_{\mu}
\label{eq:betaeq}
\end{align}
Given the mean fields and Fermi momenta determined by the coupled system of Eqs.~\rf{eq:fields} and \rf{eq:betaeq}, the total 
energy density and pressure of uniform matter follow from the energy-momentum tensor of the RMF theory. They can be written as \cite{Thakur:2025axg}
\begin{align}
\varepsilon = &\sum_{i= n,p}\frac{1}{\pi^2}\!\int_0^{k_{F,i}}\!dk\,k^2\sqrt{k^2+m^{\star 2}}
+ \frac{1}{2}m_\sigma^2\sigma^2 + \frac{1}{2}m_\omega^2\omega^2 \nonumber \\
&\qquad \,\,+ \frac{1}{2}m_\rho^2\rho^2 + \frac{b}{3}m(g_{\sigma}\sigma)^3 + \frac{c}{4}(g_{\sigma}\sigma)^4 \label{eq:energy}\\
&\qquad+ \frac{\xi}{8}g_{\omega}^4\omega^4 + 3\Lambda_\omega g_{\rho}^2 g_{\omega}^2 \rho^2 \omega^2
+ U_{\rm cut}(\sigma)
+ \sum_{\ell=e,\mu} \varepsilon_\ell, \nonumber 
\end{align}
\begin{align}
P = &\sum_{i= n,p} \frac{1}{3\pi^2}\!\int_0^{k_{F,i}}\!dk\,\frac{k^4}{\sqrt{k^2+m^{\star 2}}}
- \frac{1}{2}m_\sigma^2\sigma^2 + \frac{1}{2}m_\omega^2\omega^2 \nonumber\\
&\qquad\,\,+ \frac{1}{2}m_\rho^2\rho^2 - \frac{b}{3}m(g_{\sigma}\sigma)^3 - \frac{c}{4}(g_{\sigma}\sigma)^4 \label{eq:pressure}\\
&\qquad+ \frac{\xi}{4!}g_{\omega}^4\omega^4  + \Lambda_\omega g_{\rho}^2 g_{\omega}^2 \rho^2 \omega^2 
- U_{\rm cut}(\sigma)
+ \sum_{\ell=e,\mu} P_\ell. \nonumber 
\end{align}
The first terms in Eqs.~\rf{eq:energy} and \rf{eq:pressure} represent the kinetic contributions of nucleons 
with effective masses $m^\star$, while the subsequent terms arise from the meson fields and their self-interactions. 
Note that $U_{\rm cut}(\sigma)$ enters with opposite sign in $\varepsilon$ and $P$, as required by thermodynamic consistency. 
Leptons are treated as free relativistic Fermi gases with the usual expressions for $\varepsilon_\ell$ and $P_\ell$.

The EoS $P(\varepsilon)$ obtained from Eqs.~\rf{eq:energy} and \rf{eq:pressure} fully determines the macroscopic structure 
of nonrotating stars in General Relativity and enters directly into the radial perturbation equations discussed later.

\section{Radial oscillations of pulsating stars}

Let us consider a spherically symmetric system described by
$$ ds^2 = -e^{\nu(r)} dt^2 + e^{\lambda(r)} dr^2 + r^2 d\Omega^2$$, where the metric potentials $\nu(r)$ and $\lambda(r)$ are known functions from the Tolman–Oppenheimer–Volkoff equations. The energy density $\rho(r)$ and pressure $p(r)$ correspond to the equilibrium stellar configuration obtained from the EoS at hand. With only radial motion, Einstein's field equations may be used to compute the radial oscillation properties for a static equilibrium structure \cite{Chandrasekhar:1964zz, Chandrasekhar:1964zza, Kokkotas:2000up}. Considering the radial displacement, $\Delta r$, with the pressure perturbation as $\Delta P$, the dimensionless quantities $\xi$ = $\Delta r/r$ and $\eta$ = $\Delta P/P$ are found to satisfy the following system of coupled differential equations \cite{chanmugam1, chanmugam2}
\begin{equation}\label{ksi}
     \xi'(r) = -\frac{1}{r} \Biggl( 3\xi + \frac{\eta}{\Gamma} \Biggr) - \frac{P'(r)}{P+\rho} \xi(r),
\end{equation}
\begin{equation}\label{eta}
    \begin{split}
          \eta'(r) = \xi \Biggl[ \omega^{2} r (1+\rho/P) e^{\lambda - \nu } -\frac{4P'(r)}{P} -8\pi (P+\rho) re^{\lambda} \\
     +  \frac{r(P'(r))^{2}}{P(P+\rho)}\Biggr] + \eta \Biggl[ -\frac{\rho P'(r)}{P(P+\rho)} -4\pi (P+\rho) re^{\lambda}\Biggr] , 
    \end{split}
\end{equation} 
where $\omega$ is the frequency oscillation mode, while $\Gamma$ is the adiabatic relativistic index defined by
\begin{equation}
     \Gamma = \Biggl(1+\frac{\rho}{P} \Biggr) c_s^{2},
 \end{equation}
with $c_s^{2}$ being the speed of sound squared given by
\begin{equation}\label{cs}
     c_s^{2} = \Biggl(\frac{dP}{d\rho}\Biggr).
\end{equation}

Those two coupled differential equations, Eqs.~\eqref{ksi} and \eqref{eta}, are supplemented with two boundary conditions: one at the center, where $r$ = 0, and another at the surface, where $r$ = $R$. The boundary condition at the center requires that
\begin{equation}
    \eta = -3\Gamma \xi 
\end{equation}
must be satisfied. The equation Eq.~\eqref{eta} must be finite at the surface and hence
\begin{equation}
    \eta = \xi \Biggl[ -4 +(1-2M/R)^{-1} \Biggl( -\frac{M}{R} -\frac{\omega^{2} R^{3}}{M}\Biggr)\Biggr]
\end{equation}
must be satisfied, where $M$ and $R$ correspond to the mass and radius of the star, respectively. The frequencies are computed by
\begin{equation}
\nu = \frac{\omega}{2\pi} = \frac{s \: \omega_0}{2\pi} ~~(\text{kHz}),
\end{equation}
where $s$ is a dimensionless number, while $\omega_0 \equiv \sqrt{M/R^3}$. All oscillation equations are solved in geometric units $G=c=1$. Physical frequencies in kHz are obtained using
$$\nu(\mathrm{kHz}) = \frac{1}{2\pi} \omega \left( \frac{c^3}{G M_\odot} \right) \times 10^{-3},$$

where $\omega$ is the dimensionless eigenfrequency and $M_\odot$ is the solar mass.

\smallskip 

We use the shooting method analysis, where one starts the integration for a trial value of $\omega^2$ and a given set of initial values that satisfy the boundary condition at the center. We integrate towards the surface, and the discrete values of $\omega^2$ for which the boundary conditions are satisfied correspond to the eigenfrequencies of the
radial perturbations.

\smallskip

Those equations represent the Sturm-Liouville eigenvalue equations for $\omega$. The solutions provide the discrete eigenvalues $\omega_n^{2}$ and can be ordered as 
\begin{equation*}
\omega_0 ^{2} < \omega_1 ^{2} <... <\omega_n ^{2}, 
\end{equation*}
where $n$ is the number of nodes for a star of a given stellar mass and radius. Finally, once the spectrum is known, the so-called large frequency separation may be computed as follows
\begin{equation}
\Delta \nu_n = \nu_{n+1} - \nu_n, \; \; \; n=0,1,2,3,...
\end{equation}
which is the difference between consecutive modes, and which is widely used in Asteroseismology \cite{tassoul, Chaplin:2013hha, ilidio}.


\section{Results and Discussion}

\begin{figure}
\centering
\includegraphics[width=0.8\linewidth]{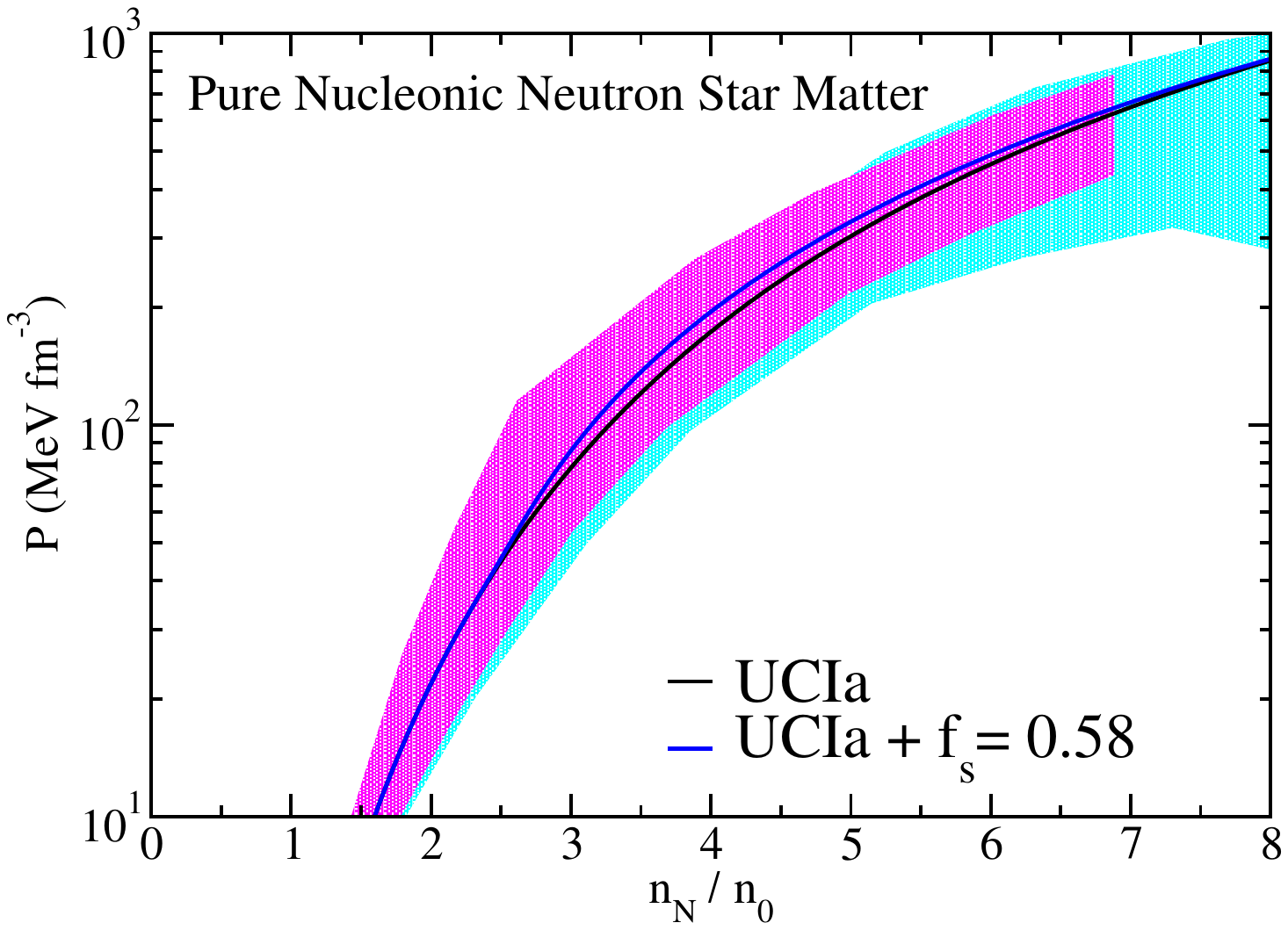}
\caption{
The pressure of pure nucleonic neutron star matter as a
function of nucleon number density $n_N$ (in units of $n_0$) is shown for
the original UCIa and UCIa with $f_{s}$= 0.58. Additional
constraints (as mentioned in the text) are represented by shaded
regions.
}
\label{fig:1}
\end{figure}

In this section, we present our numerical results for microscopic and macroscopic observables of a pure nucleonic neutron star.
To explore the impact of the $\sigma$-cutoff potential on both microscopic and macroscopic observables, we employ the calibrated original UCIa and UCIa with $f_{s}$=0.58 RMF parameterizations as baselines. We illustrate the pressure
of pure nucleonic neutron star matter as a function of nuclear number density $n_{N}$ (in units of $n_{0}$) for 
original UCIa and UCIa with $f_{s}$= 0.58 in Figure \ref{fig:1}. The NICER collaboration \cite{Riley:2019,Miller:2019} made significant
progress in precise radius measurements by simultaneously determining the radius and mass of PSR J0030+0451 through 
x-ray pulse-profile modeling. The impact of this measurement on the equation of state was studied using two different 
parametrizations: a speed of sound (CS) model \cite{Greif:2019} and a piecewise polytropic (PP) model \cite{Hebeler:2013},
as detailed in Ref. \cite{Raaijimakers:2019}. A combined analysis of these models was performed to infer the implications on the EoS 
based on the NICER measurement, GW170817, and the 2.14$M_{\odot}$ , as described in Ref. \cite{Raaijimakers:2020} 
and represented as a cyan band (CS model) and magenta band (PP model) in Figure \ref{fig:1}. Figure \ref{fig:1} clearly indicates that
the EoS of original UCIa and UCIa with $f_{s}$= 0.58 are consistent with the CS and PP models. 
The effect of the $\sigma$-cutoff potential starts to appear at 2.65$n_{0}$, where the EoS of UCIa with $f_{s}$= 0.58 
overlaps with original UCIa upto 2.65$n_{0}$, and beyond that point, the EoS of UCIa with $f_{s}$= 0.58 becomes stiffer 
compared to the original UCIa parameter set. 

\begin{figure}
\centering
\includegraphics[width=0.8\linewidth]{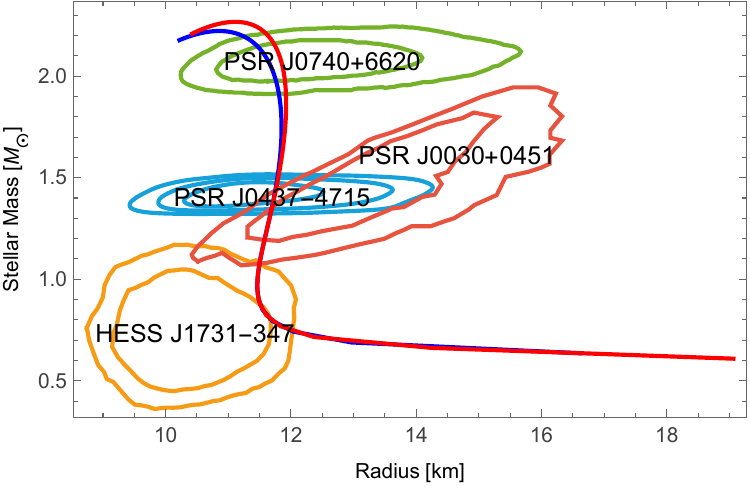}
\caption{
Mass-to-radius relationships for the two viable equations-of-state considered in this work (original UCIa and UCIa with $f_s$=0.58). They are in agreement with current astrophysical constraints from NICER results \cite{Riley:2019,Miller:2019,
Choudhury:2024xbk,Miller:2021qha,Riley:2021pdl} and the HESS J1731-347 compact object \cite{Doroshenko:2022nwp}. Furthermore, they are capable of accommodating massive stars at two solar masses
\cite{Demorest:2010,Antoniadis:2013,Cromartie:2019}, while at the same time satisfy the stringent constraints for the 
canonical star at 1.4 solar masses obtained in \cite{Capano:2019eae}.
}
\label{fig:2}
\end{figure}

In Fig.~\ref{fig:2} we have displayed the mass-to-radius relationships for the two viable equations-of-state considered in this work (original UCIa and UCIa with $f_s$=0.58). They are in agreement with current astrophysical constraints from NICER results \cite{Riley:2019,Miller:2019,
Choudhury:2024xbk,Miller:2021qha,Riley:2021pdl} and the HESS J1731-347 compact object \cite{Doroshenko:2022nwp}. Furthermore, they are capable of accommodating massive stars at two solar masses
\cite{Demorest:2010,Antoniadis:2013,Cromartie:2019}, while at the same time satisfy the stringent constraints for the 
canonical star at 1.4 solar masses obtained in \cite{Capano:2019eae}.

\begin{figure}
\centering
\includegraphics[width=0.8\linewidth]{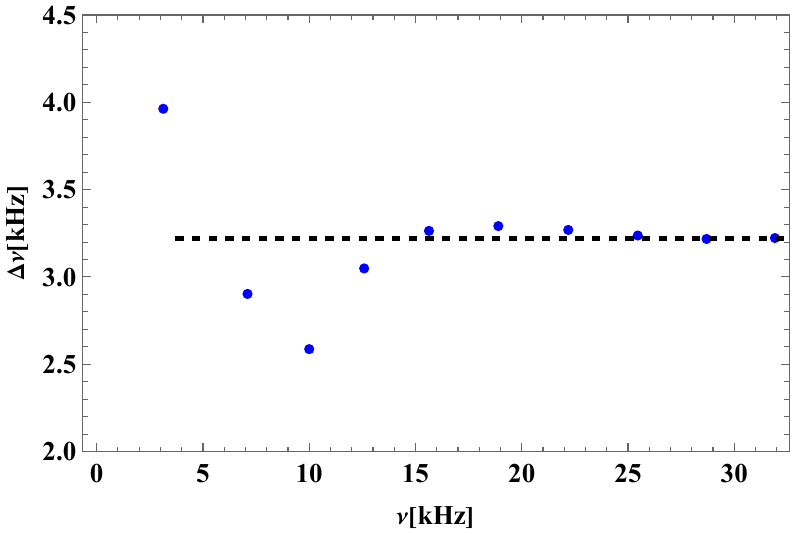} \
\includegraphics[width=0.8\linewidth]{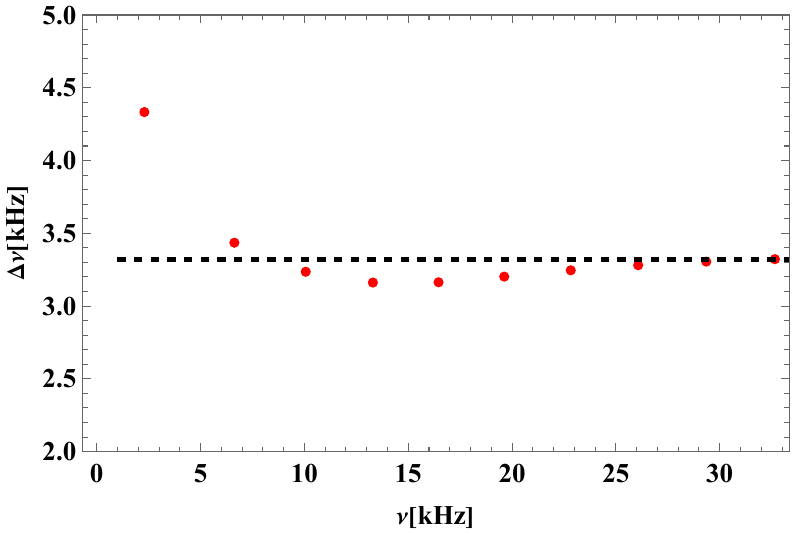} 
\caption{
Large frequency separations for the purely nucleonic EoS without cuts.
{\bf Top:} Spectrum considering $M=1.4~M_{\odot}$ (blue color).
{\bf Bottom:} Spectrum considering $M=2.0~M_{\odot}$ (red color).
}
\label{fig:3}
\end{figure}

\begin{figure}
\centering
\includegraphics[width=0.8\linewidth]{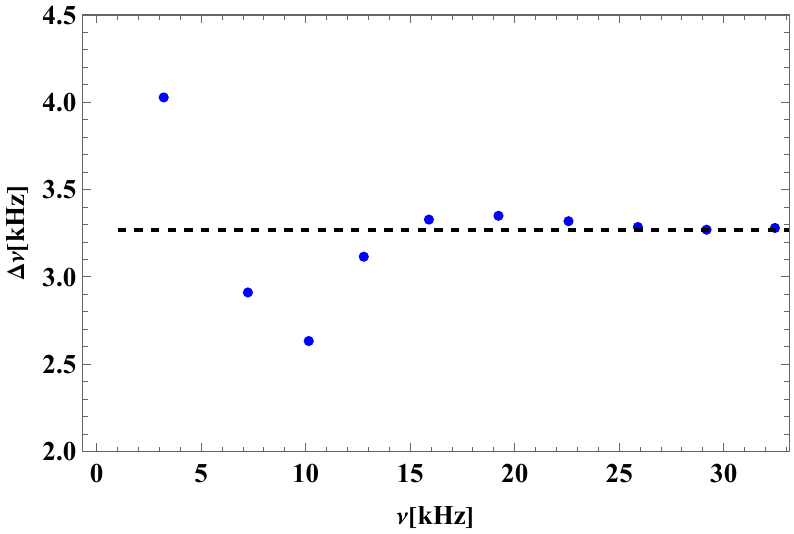} \
\includegraphics[width=0.8\linewidth]{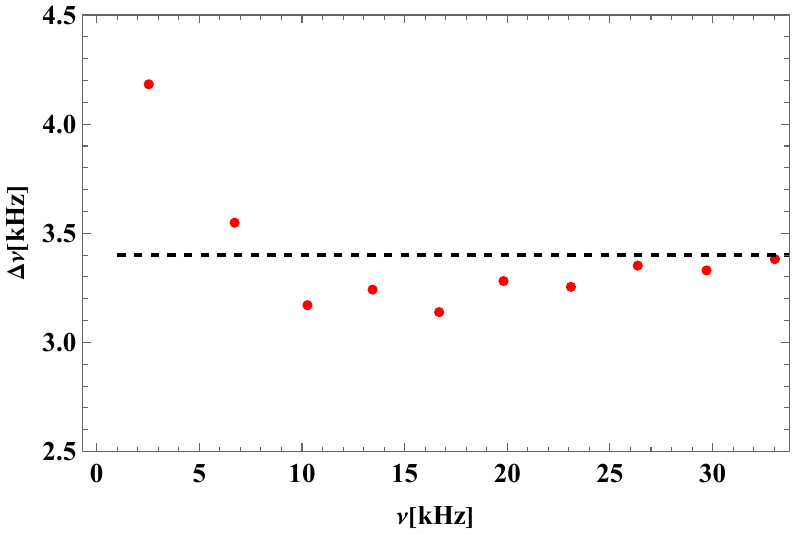} 
\caption{
Large frequency separations for the purely nucleonic EoS with $f_s=0.58$.
{\bf Top:} Spectrum considering $M=1.4~M_{\odot}$ (blue color).
{\bf Bottom:} Spectrum considering $M=2.0~M_{\odot}$ (red color).
}
\label{fig:4}
\end{figure}

\begin{figure}
\centering
\includegraphics[width=0.8\linewidth]{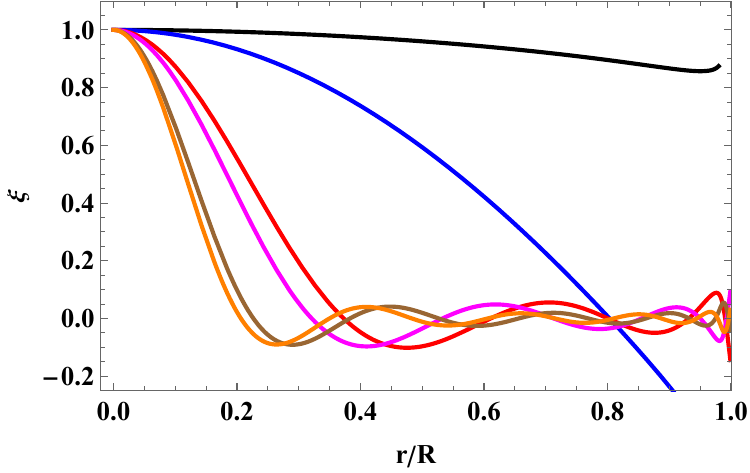} \
\includegraphics[width=0.8\linewidth]{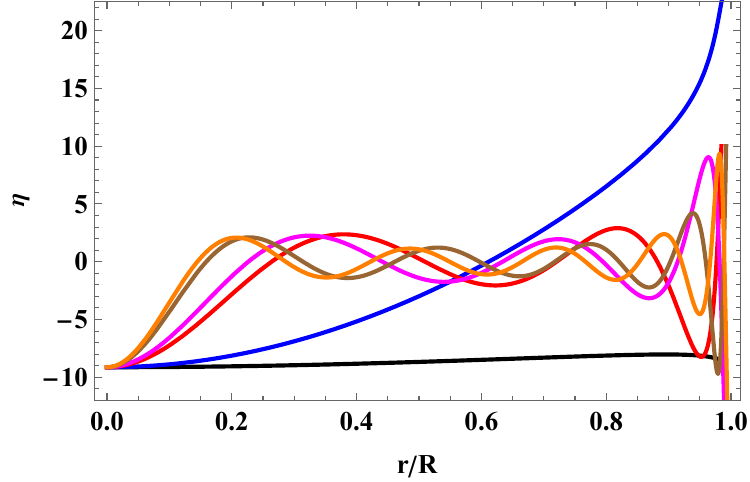} 
\caption{
Radial profiles of the eigenfunctions $\xi(r), \eta(r)$ versus dimensionless radial coordinate for the case $f_s=0, M=1.4~M_{\odot}$.
Shown are the fundamental and first excited mode ($n=0,1$) in black and blue, intermediate excited modes ($n=5,6$) in red and magenta as well as highly excited modes ($n=9,10$) in brown and orange.
}
\label{fig:5}
\end{figure}

\begin{figure}
\centering
\includegraphics[width=0.8\linewidth]{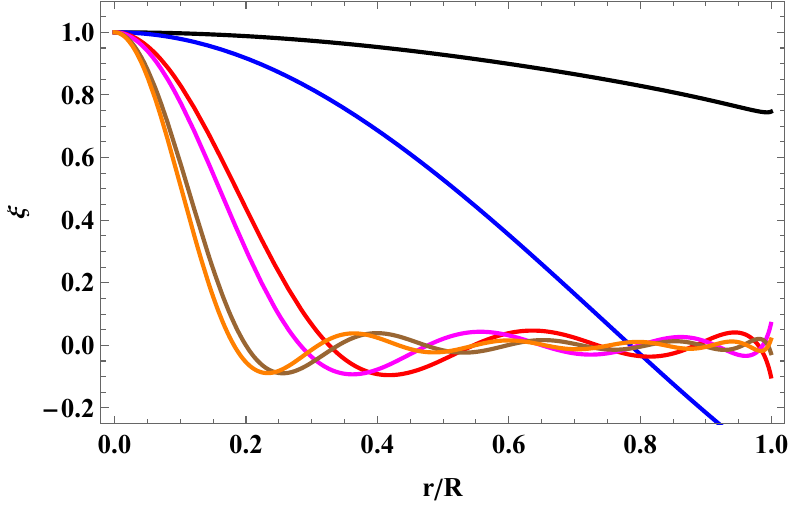} \
\includegraphics[width=0.8\linewidth]{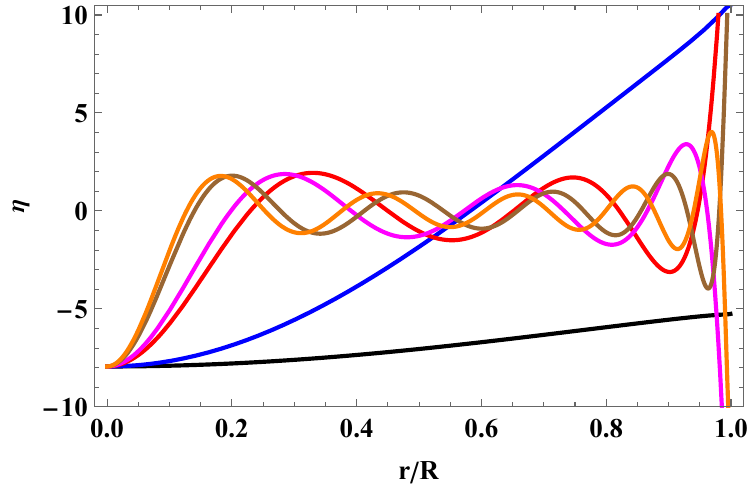} 
\caption{
Radial profiles of the eigenfunctions $\xi(r), \eta(r)$ versus dimensionless radial coordinate for the case $f_s=0, M=2.0~M_{\odot}$.
Shown are the fundamental and first excited mode ($n=0,1$) in black and blue, intermediate excited modes ($n=5,6$) in red and magenta as well as highly excited modes ($n=9,10$) in brown and orange.
}
\label{fig:6}
\end{figure}

\begin{figure}
\centering
\includegraphics[width=0.8\linewidth]{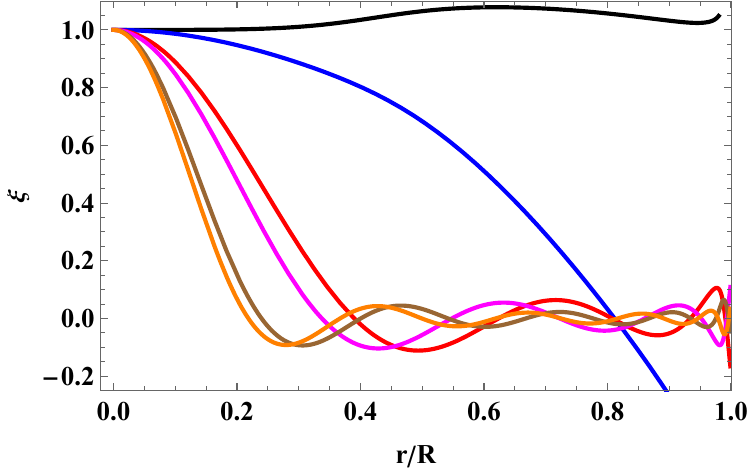} \
\includegraphics[width=0.8\linewidth]{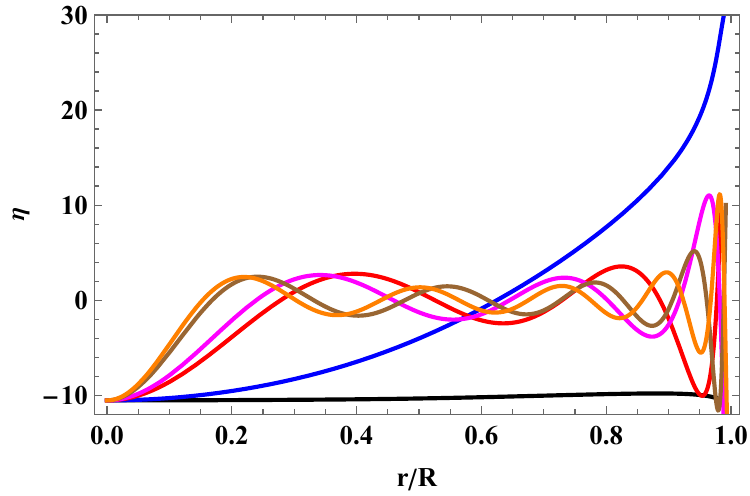} 
\caption{
Radial profiles of the eigenfunctions $\xi(r), \eta(r)$ versus dimensionless radial coordinate for the case $f_s=0.58, M=1.4~M_{\odot}$.
Shown are the fundamental and first excited mode ($n=0,1$) in black and blue, intermediate excited modes ($n=5,6$) in red and magenta as well as highly excited modes ($n=9,10$) in brown and orange.
}
\label{fig:7}
\end{figure}

\begin{figure}
\centering
\includegraphics[width=0.8\linewidth]{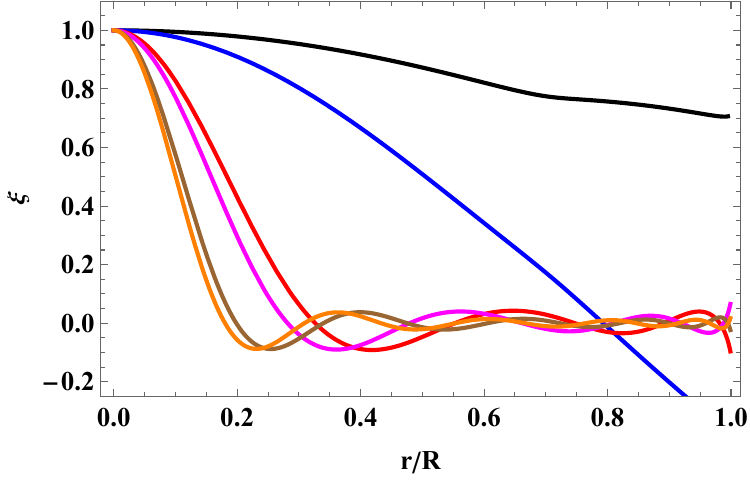} \
\includegraphics[width=0.8\linewidth]{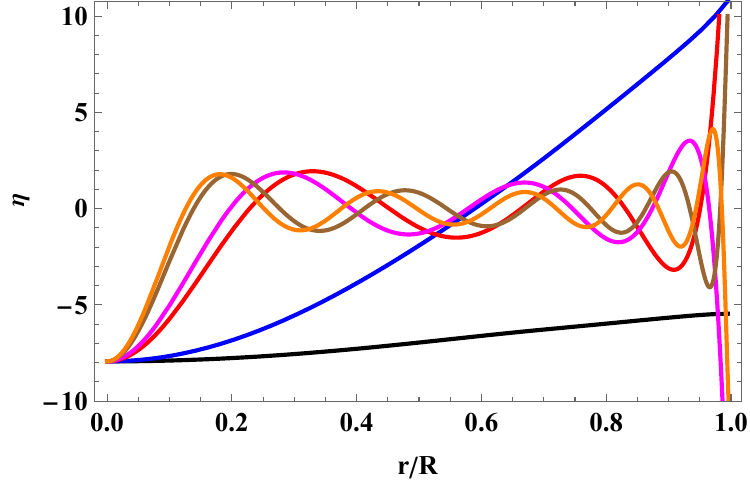} 
\caption{
Radial profiles of the eigenfunctions $\xi(r), \eta(r)$ versus dimensionless radial coordinate for the case $f_s=0.58, M=2.0~M_{\odot}$.
Shown are the fundamental and first excited mode ($n=0,1$) in black and blue, intermediate excited modes ($n=5,6$) in red and magenta as well as highly excited modes ($n=9,10$) in brown and orange.
}
\label{fig:8}
\end{figure}

As far as radial oscillations of pulsating stars are concerned, the values of the frequencies are shown in Table~\ref{tab:1} assuming two stellar masses, namely $M=1.4~M_{\odot}, M=2.0~M_{\odot}$, for each EoS considered in this work. The large difference separations for the original UCIa are shown in Fig.~\ref{fig:3} and for UCIa with $f_s$=0.58 are shown in Fig.~\ref{fig:4}. Finally, the radial profiles of the perturbations $\xi(r),\eta(r)$ for all four cases are shown in Figures \ref{fig:5}, \ref{fig:6}, \ref{fig:7} and \ref{fig:8}.



\begin{table}[!t]
\centering
\fontsize{10pt}{14.4pt}\selectfont
\setlength{\tabcolsep}{3.5pt}
\renewcommand{\arraystretch}{1.05}
\begin{tabular}{|c|c|c|c|c|}
\hline
\multirow{2}{*}{\diagbox[dir=SW]{Mode $n$}{Mass}}
& \multicolumn{2}{c|}{PN0}
& \multicolumn{2}{c|}{PN0.58} \\
\cline{2-5}
& $1.4\,M_{\odot}$ & $2.0\,M_{\odot}$
& $1.4\,M_{\odot}$ & $2.0\,M_{\odot}$ \\
\hline\hline
0  & 3.13   & 2.30  & 3.20  & 2.54   \\ \hline
1  & 7.10   & 6.63  & 7.23  &  6.72  \\ \hline
2  & 10.00  & 10.06 & 10.14 &  10.27  \\ \hline
3  & 12.59  & 13.30 & 12.78 & 13.44   \\ \hline
4  & 15.63  & 16.46 & 15.89 &  16.68  \\ \hline
5  & 18.90  & 19.62 & 19.22 &  19.82  \\ \hline
6  & 22.19  & 22.83 & 22.57 &  23.10  \\ \hline
7  & 25.46  & 26.07 & 25.89 &  26.35  \\ \hline
8  & 28.70  & 29.35 & 29.18 &  29.71  \\ \hline
9  & 31.91  & 32.66 & 32.45 &  33.04  \\ \hline
10 & 35.14  & 35.98 & 35.73 &  36.42  \\ \hline
\end{tabular}%

\vspace{1mm}

\caption{Frequencies (in kHz) of radial oscillation modes for the four cases discussed here, see text.}
\label{tab:1}
\end{table}



\section{Summary and concluding remarks}\label{sec7} 

We studied microscopic equations of state for cold, $\beta$-equilibrated \emph{pure nucleonic} matter within RMF theory, using UCIa as the baseline \cite{Malik:2024} and supplementing it with a high-density $\sigma$-cutoff regulator $U_{\rm cut}(\sigma)$ \cite{Maslov:2015,Patra:2022}. The role of the cutoff is to curb the growth of the scalar $\sigma$ field at supranuclear density, thereby suppressing scalar attraction (increasing the effective baryon masses) and stiffening the EoS primarily in the high-density sector, while leaving the near-saturation calibration essentially intact \cite{Maslov:2015,Patra:2022}. By comparing the representative cases $f_s=0$ and $f_s=0.58$, we quantified how this controlled scalar-sector modification propagates from microphysics to global structure and to dynamical stability.

At the macroscopic level, we solved the Tolman--Oppenheimer--Volkoff equations and the tidal perturbation equations to obtain mass--radius relations and tidal deformabilities for nonrotating neutron stars. The $\sigma$-cutoff stiffening increases the maximum supported mass and shifts radii and tidal deformabilities accordingly, enabling $2M_\odot$ configurations while remaining compatible with current constraints on $R_{1.4}$ and $\Lambda_{1.4}$ inferred from X-ray timing and gravitational-wave observations \cite{Demorest:2010,Antoniadis:2013,Cromartie:2019,Riley:2019,Miller:2019,Abbott:2018,Abbott:2019}.

A central result of this work is that radial oscillations provide an additional dynamical consistency check beyond static and tidal constraints. We computed the frequencies and eigenfunctions of 11 radial oscillation modes (the fundamental mode and 10 excited modes) and used the sign change of $\omega_0^2$ to identify the onset of radial instability \cite{Chandrasekhar:1964zz,Kokkotas:2000up}. For a given stellar mass, the stiffened model with $f_s=0.58$ yields systematically higher eigenfrequencies than the baseline case, reflecting the increased pressure support at supranuclear density. At higher overtones, the large frequency separations tend towards an approximately constant value, consistent with asymptotic expectations \cite{tassoul}. The corresponding eigenfunctions exhibit the standard nodal structure: the number of nodes increases with overtone number, with zero nodes for the fundamental mode, one node for the first excited mode, and so on. Taken together, these mode systematics provide a stringent cross-check: an EoS variant that satisfies multimessenger constraints must also yield a radially stable sequence (i.e. $\omega_0^2>0$) up to the maximum mass.

The framework developed here admits several concrete extensions. Incorporating rotation would allow one to assess how spin modifies both equilibrium sequences and mode spectra, and to test the robustness of the stability conclusions in a more realistic setting \cite{Paschalidis:2016vmz}. Extending the microphysics to finite temperature and, where relevant, lepton-rich conditions would connect the present cold-star analysis to proto-neutron stars and merger remnants. Finally, repeating the same programme in alternative theories of gravity would enable a controlled comparison between EoS-driven and gravity-driven modifications of stellar structure and radial dynamics \cite{Staykov:2014mwa}. These directions would broaden the use of radial oscillations as a dynamical consistency test alongside multimessenger constraints. 

\acknowledgements

Y.~K. would like to acknowledge that this work was supported by the Universidad Nacional Autónoma de México Postdoctoral Program (POSDOC). A. \"O. would like to acknowledge networking support of the COST Action CA21106 - COSMIC WISPers in the Dark Universe: Theory, astrophysics and experiments (CosmicWISPers), the COST Action CA22113 - Fundamental challenges in theoretical physics (THEORY-CHALLENGES), the COST Action CA21136 - Addressing observational tensions in cosmology with systematics and fundamental physics (CosmoVerse), the COST Action CA23130 - Bridging high and low energies in search of quantum gravity (BridgeQG), and the COST Action CA23115 - Relativistic Quantum Information (RQI) funded by COST (European Cooperation in Science and Technology).



\end{document}